\documentclass[12pt]{article}
\usepackage{graphicx} 
\usepackage{graphics,cite,amssymb,epsfig,float,psfrag}
\usepackage[usenames,dvips]{color}
\usepackage{rotating}

\usepackage{epsfig,amssymb}
\usepackage{latexsym}

\RequirePackage{lineno}

\newcommand{\bq}{\begin{equation}}
\newcommand{\eq}{\end{equation}}
\newcommand{\bqa}{\begin{eqnarray}}
\newcommand{\eqa}{\end{eqnarray}}
\newcommand{\ben}{\begin{enumerate}}
\newcommand{\een}{\end{enumerate}}
\newcommand{\bc}{\begin{center}}
\newcommand{\ec}{\end{center}}
\newcommand{\bqb}{\begin{eqnarray*}}
\newcommand{\eqb}{\end{eqnarray*}}

\newcommand{\qsl}{q\hskip-0.21cm\slash}
\newcommand{\qpsl}{q'\hskip-0.29cm\slash}
\newcommand{\esl}{e\hskip-0.21cm\slash}
\newcommand{\eesl}{\epsilon\hskip-0.21cm\slash}

\newcommand{\beq}{\begin{eqnarray}} 
\newcommand{\eeq}{\end{eqnarray}} 

\begin{document}


\vspace{1cm}

\vspace*{.5cm}

\begin{center}

{\large\bf  The relevance of polarized bZ production at LHC }

\vspace*{.8cm}

\mbox{\large  M. Beccaria$^1$, N. Orlando$^1$, G. Panizzo$^2$, F.M. Renard$^3$ and C. Verzegnassi$^2$}

\vspace*{.8cm}

$^1$ Dipartimento di Matematica e Fisica ``Ennio De Giorgi'', Universit\`a del Salento and  INFN, Sezione di Lecce,
Italy. \\
$^2$ Dipartimento di Fisica , Universit\`a di Trieste and INFN Sezione di 
Trieste, Italy. \\
$^3$Laboratoire Univers et Particules de Montpellier, UMR
5299. Universit\'e Montpellier II, Place Eug\`ene Bataillon CC072 F-34095 Montpellier Cedex 5.\\

\end{center}

\vspace{1.4cm}

\begin{abstract} 

We consider the Z polarization asymmetry $A_{Z}=(\sigma(Z_{R})-\sigma(Z_{L}))/(\sigma(Z_{R})+\sigma(Z_{L}))$
in the process of associated bZ production at the LHC. We show that in the Standard Model (SM) this quantity is essentially given by its Born approximation, 
remaining almost unaffected by QCD scales and parton
distribution functions variations as well as by electroweak corrections. The theoretical quantity that appears in $A_{Z}$ is the 
same that provides the LEP1 $Z\rightarrow b\overline{b}$ 
forward-backward asymmetry, the only measured observable still in some contradiction with the SM prediction. In this sense, 
$A_{Z}$ would provide the possibility of an independent verification of the possible SM discrepancy, which could reach, if consistency with LEP1 measurements is imposed, 
values of the relative ten percent size.

\end{abstract} 

\thispagestyle{empty}
\setcounter{page}{0}
\newpage

\subsection*{1. Introduction} 

The Standard Model is confirmed up to
per-mille precision by collider data~\cite{EWFit}; moreover,
very recently, Higgs boson signals~\cite{ATLASHiggs,CMSHiggs}~seems to rise in a narrow mass
window around 125 GeV, consistently with predictions based on global fits to electroweak data~\cite{EWFit}.

The question arises of whether all the theoretical SM predictions have been confirmed by the related experimental measurements. 
The answer to this fundamental question is nowadays that  at least one
experimental result still appears in some sizable contradiction
(roughly, at $3\, \sigma$ level) with the SM , i.e. the
measurement of the forward-backward asymmetry of $b\overline{b}$
production at the Z peak~\cite{LEPEWWG}, $A_{FB}^{b}$.

In fact, a number of models have been proposed that might cure the
discrepancy (see~\cite{ZbbBSM1,ZbbBSM2,ZbbBSM3} and references therein). 
In particular, a slightly embarrassing fact for Supersymmetric models is the difficulty that the simplest MSSM version would face to eliminate the discrepancy, 
as exhaustively discussed in Ref.~\cite{ZbbMSSM}. 

The aim of this paper is that of showing that a specific observable can 
be defined at LHC that would provide essentially a re-measurement of the same LEP1 $A_{FB}^{b}$ quantity, in spite of the total difference of the produced final state. 
This quantity is defined in the production of a bZ pair as the ratio
of the difference of production cross sections with different (left,
right) Z polarizations ($Z_L$, $Z_R$)  divided by the corresponding sum. 

We shall first show in section 2 that this quantity is straightforwardly proportional to $A_{FB}^{b}$ at the simplest partonic Born level, 
providing a possible ten percent deviation from its SM prediction if the relevant parameters are chosen to reproduce the experimental LEP1 result for the asymmetry.
In Section 3 we shall derive the special property of our considered quantity, i.e. the fact that it remains unaffected, at realistic levels, by variations of the strong scales and of the adopted 
parton distribution functions (pdfs) as well as by electroweak
corrections. This would represent, in our opinion, a strong motivation to perform an accurate measurement of the asymmetry at LHC in a not far future, as qualitatively discussed in the final conclusions.

\subsection*{2. The Z polarization asymmetry at tree--level}

\begin{figure}[!]
\vspace{1.5cm}
\begin{center}
\begin{tabular}{cc}
\includegraphics[width=0.4\textwidth]{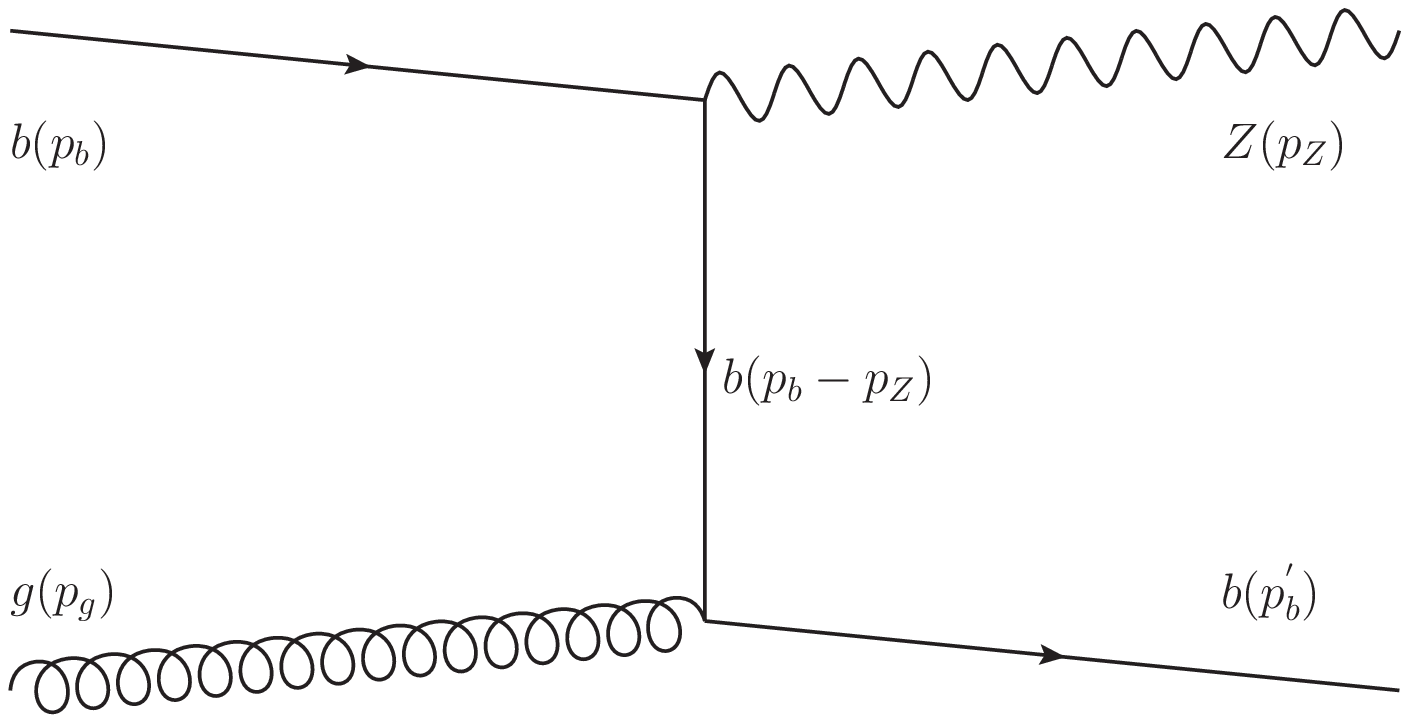}&
\includegraphics[width=0.38\textwidth]{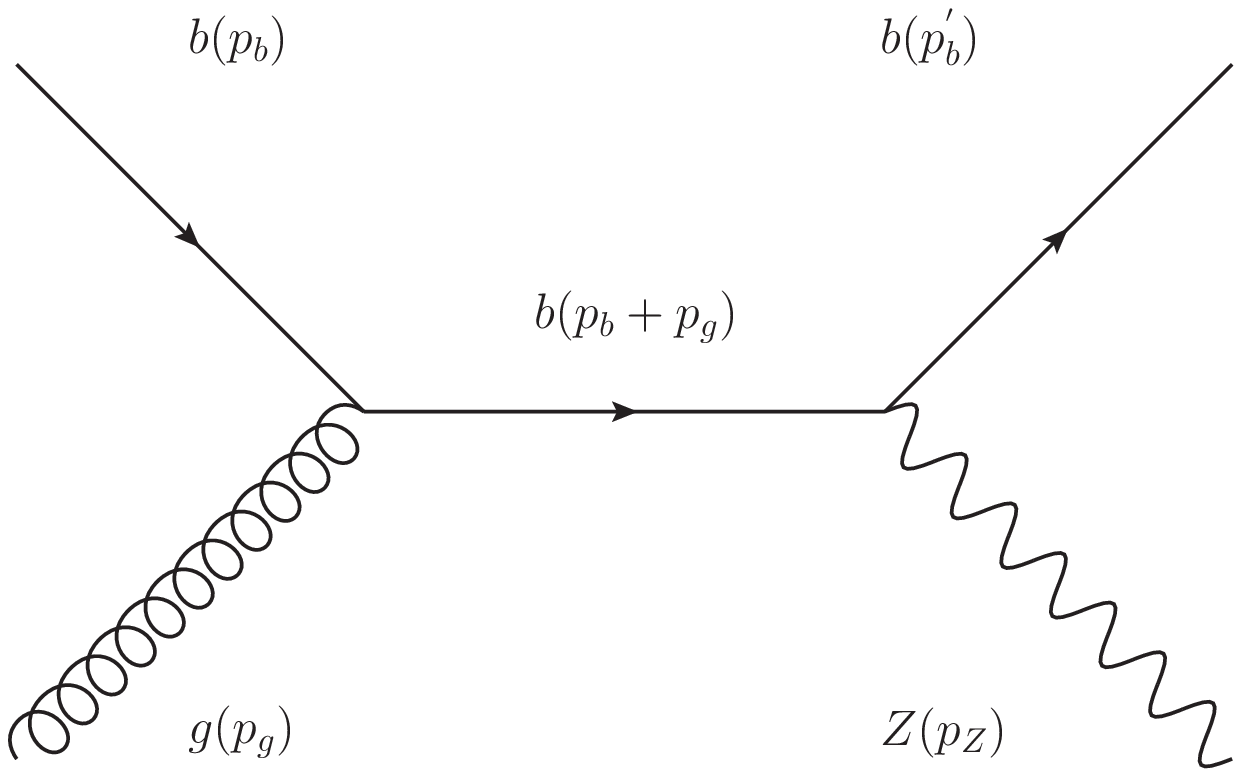}\\
\end{tabular}
\caption[]{Born diagrams for associated production of a Z boson and a single b-quark.}
\label{fig:diag}
\end{center}
\end{figure}

We shall consider the process of associated production of a Z boson 
and a single b-quark, represented in Figure~\ref{fig:diag},
defined at parton level by subprocesses $bg\to bZ$ involving two Born 
diagrams with bottom quark exchange in the $s$-channel and in the $u$-channel. 

This process has been calculated at next-to-leading order in QCD in a previous
paper~\cite{ZbNloQCD}~where the theoretical uncertainties assessment on
cross section calculation have been addressed as well. 

For our purposes we need, though, a derivation of the expressions of the polarized cross sections. 
This requires a number of formulae that we shall briefly show in what follows, starting from the calculation of the various quantities performed at the Born level.

The interaction vertices involved in the diagrams of Figure~\ref{fig:diag}~are defined as follow 

\bq
gqq:~~ig_s\esl \left({\lambda^l\over2} \right )~~~~~~~~
Zbb: -ie\eesl (g^L_{Zb}P_L+g^R_{Zb}P_R) ~~ ,
\eq

Therefore, the Born invariant amplitude is given by

\bqa
A^{Born}(gb\to Zb)&=&eg_s\left({\lambda^l\over2}\right)
\bar u(b')~(~\eesl (g^L_{Zb}P_L+g^R_{Zb}P_R)
{(\qsl+m_b)\esl \over s-m^2_b}\nonumber\\
&&
+{\esl (\qpsl+m_b)\eesl\over u-m^2_b}
(g^L_{Zb}P_L+g^R_{Zb}P_R)~)~u(b) ~~ ,
\eqa
\noindent
where $e$, $\lambda^l$ are the gluon polarization vector and
colour matrix, $\epsilon$ is the $Z$ polarization vector
and $q=p_b+p_g=p_Z+p'_b$, $s=q^2$, $q'=p'_b-p_g=p_b-p_Z$,
$u=q^{'2}$ with the kinematical decompositions 

\bq
p_b=(E_b;0,0,p)~~  , ~~~~
p'_b=(E'_b;p'\sin\theta,0,p'\cos\theta) ~~ ,
\eq
\bq
p_g=(p;0,0,-p)~~ , ~~~~ p_Z=(E_Z;-p'\sin\theta,0,-p'\cos\theta) ~~ ,
\eq
\bq
e(g)=\left (0;{\mu\over\sqrt{2}},-~{i\over\sqrt{2}},0 \right )~~ , ~~~~
\epsilon(Z_T)=\left (0;{\mu'\cos\theta\over\sqrt{2}},{i\over\sqrt{2}},
{-\mu'\sin\theta\over\sqrt{2}} \right ) ~~ ,
\eq
\bq
\epsilon(Z_0)=\left (-{p'\over M_Z};{E_Z\over M_Z}\sin\theta,0,
{E_Z\over M_Z}\cos\theta \right ) ~~ .
\eq
            
The decomposition of Dirac spinors and polarization vectors leads to 24 helicity amplitudes
denoted as $F_{\lambda \mu\tau\mu'}$
with $\lambda=\pm{1\over2}$, $\mu=\pm1$,
$\tau\pm{1\over2}$, $\mu'=\pm1,0$ referring to $b,g,b',Z$ respectively.

However, in order to explore quickly by hand the properties of this subprocess
we can neglect $m_b/M_Z$ and $m_b/\sqrt{s}$ terms  and consider only the following
eight non vanishing amplitudes: six transverse ones

\bq
 F_{++++}={2eg_sg^R_{Zb}\sqrt{\beta'}\over\cos{\theta\over2}}   
~~ , ~~~~ F_{----}={2eg_sg^L_{Zb}\sqrt{\beta'}\over\cos{\theta\over2}}
 ~~ ,\eq

\bq
F_{+-+-}=2eg_sg^R_{Zb}{\cos{\theta\over2}\over\sqrt{\beta'}}
~~ , ~~~~ F_{-+-+}=2eg_sg^L_{Zb}{\cos{\theta\over2}\over\sqrt{\beta'}}
~~ , \eq

\bq
 F_{+-++}=2eg_sg^R_{Zb}\cos{\theta\over2}.{M^2_Z\over s}.   
{\tan^2{\theta\over2}\over\sqrt{\beta'}}
 ~~ , ~~~~ F_{-+--}=2eg_sg^L_{Zb}\cos{\theta\over2}.{M^2_Z\over s}.   
{\tan^2{\theta\over2}\over\sqrt{\beta'}} ~~ , \eq

and two longitudinal ones 

\bq
 F_{+-+0}=-2\sqrt{2}eg_sg^R_{Zb}{\sin{\theta\over2}\over\sqrt{\beta'}}.{M_Z\over \sqrt{s}}   
~~ , \eq

\bq
 F_{-+-0}=2\sqrt{2}eg_sg^L_{Zb}{\sin{\theta\over2}\over\sqrt{\beta'}}.{M_Z\over \sqrt{s}}   
~~ , \eq
\noindent
having defined
\bq
\beta'={2p'\over\sqrt{s}}\simeq 1-{M^2_Z\over s}
~~ . \eq

For our analysis it is instructive to consider the $Z$ density matrix

\bq
\rho^{ij}=\sum_{\lambda \mu\tau}
F_{\lambda \mu\tau i}F^*_{\lambda \mu\tau j}
~~ . \eq

A priori there are nine independent $Z$ density matrix elements.
However with the above Born terms and neglecting again the subleading terms in $m_b$ they reduce to only five ones

\bq
\rho^{++}=4e^2g^2_s \left (
{g^{R2}_{Zb}\over\cos^2{\theta\over2}}\left (\beta'+{\sin^4{\theta\over2}\over\beta'}
\left({M^2_Z\over s}\right)^2 \right)
+g^{L2}_{Zb}{\cos^2{\theta\over2}\over\beta'} \right )
\label{rhopp} ~~ ,\eq
\bq
\rho^{--}=4e^2g^2_s \left (
{g^{L2}_{Zb}\over\cos^2{\theta\over2}}\left (\beta'+{\sin^4{\theta\over2}\over\beta'}
\left({M^2_Z\over s}\right)^2 \right )
+g^{R2}_{Zb}{\cos^2{\theta\over2}\over\beta'} \right )
\label{rhomm} ~~ , \eq
\bq
\rho^{00}=8e^2g^2_s
(g^{R2}_{Zb}+g^{L2}_{Zb})\sin^2{\theta\over2}
\left({M^2_Z\over s\beta'}\right)
\label{rho00} ~~ , \eq
\bq
\rho^{+0}=\rho^{0+}=-4e^2g^2_s
g^{R2}_{Zb}{\sin^3{\theta\over2}\over\cos{\theta\over2}}
\left({M^3_Z\sqrt{2}\over \beta's\sqrt{s}}\right)
~~ , ~~~~ 
\rho^{-0}=\rho^{0-}=4e^2g^2_s
g^{L2}_{Zb}{\sin^3{\theta\over2}\over\cos{\theta\over2}}
\left({M^3_Z\sqrt{2}\over \beta's\sqrt{s}}\right)
\label{rhop0} ~~ . \eq


With these powerful but extremely simple mathematical expressions at
hand, we can explore some physical observables of the process under
consideration that keep informations of the Zb vertex structure. Let's
stick for the moment at the partonic level. The first obvious quantity
that one can inspect is the subprocess unpolarized angular
distribution: with the colour sum  
$\sum_{col} Tr({\lambda^l\over2}{\lambda^l\over2})=4$,
the unpolarized subprocess angular distribution (averaged on gluon and $b$ spins and colours)
is given by

\bq {d\sigma\over d\cos\theta}={\beta'\over768\pi s\beta}
\sum_{\lambda \mu\tau\mu'}
|F_{\lambda \mu\tau\mu'}|^2
~~ . \eq

One sees that it is
proportional to $(\rho^{++}+\rho^{--}+\rho^{00})$ and, summing the above density matrix expressions, solely depends on
$(g^R_{Zb})^2+(g^L_{Zb})^2$:

\bq
\sum_{\lambda \mu\tau\mu'}
|F_{\lambda \mu\tau\mu'}|^2=((g^R_{Zb})^2+(g^L_{Zb})^2)C_{diff} ~~ , 
\eq

with

\bq
C_{diff}={4e^2g^2_s \beta'\cos^2{\theta\over2}}\left({1\over\cos^4{\theta\over2}}
+{1\over\beta^{'2}}+\left({M^2_Z\over s}  
{\tan^2{\theta\over2}\over\beta'}\right)^2
+{2M^2_Z\over s\beta^{'2}}\tan^2{\theta\over2}\right) ~~ .
\eq

In order to separate the $g_{Zb}^R$ and $g_{Zb}^L$ contributions, and
to check so their possible anomalous behaviors, one needs to be
sensitive to different density matrix combinations than the sum just
found in the unpolarized distribution. This can be achieved only
keeping track of the final Z polarization. The general procedure of
its measurement has been described in~\cite{Zpol1,Zpol2}~for Tevatron
processes. The Z polarization can be analyzed by looking at Z decay distributions, 
for example in lepton pairs.
It is shown that each density matrix element is associated to a specific 
$\theta_l,\phi_l$ dependence. The polarized quantities, therein called $\sigma^P$ and $\sigma^I$,
respectively proportional to 
$(\rho^{++}-\rho^{--})$ and to $(\rho^{+0}-\rho^{-0})$
are the only ones in which the combination
$(g^R_{Zb})^2-(g^L_{Zb})^2$ appears, as one can check by using the
above expressions (\ref{rhopp}-\ref{rhop0}) of the density matrix elements. 
They respectively produce lepton
angular dependencies of the types
$\cos\theta_l$ and $\sin2\theta_l\cos\phi_l$ as compared to the unpolarized part
proportional to $(1+\cos^2\theta_l)$. The specific generalization of that analysis
to the LHC case is under consideration at the moment. 

From this brief discussion, we are naturally led to define the Z boson
polarization asymmetry $A_{Z}$ in bZ production as 


\bq
A_Z\equiv {\sigma(Z_R)-\sigma(Z_L)\over \sigma(Z_R)+\sigma(Z_L)} =
{(g^R_{Zb})^2-(g^L_{Zb})^2\over (g^R_{Zb})^2+(g^L_{Zb})^2} C_{pol} 
\label{Az1} ~~ , \eq


where $C_{pol}$ is given as a convolution involving the bottom quark ($b$
and $\overline{b}$) and gluon ($g$) pdfs: 
\bq
C_{pol}= \frac { (bg+\overline{b}g)\otimes \left( {1\over\cos^4{\theta\over2}}-{1\over\beta^{'2}}
+\left({M^2_Z\over s}.   
{\tan^2{\theta\over2}\over\beta'}\right)^2\right) }
{(bg+\overline{b}g)\otimes \left( {1\over\cos^4{\theta\over2}}+{1\over\beta^{'2}}
+\left({M^2_Z\over s}.   
{\tan^2{\theta\over2}\over\beta'}\right)^2 \right)}
 ~~ . \eq
As one sees, Eq.~\ref{Az1}~is simply proportional to the asymmetry parameter $\mathcal{A}_{b}$:

\bq
\mathcal{A}_{b} = 
{(g^L_{Zb})^2-(g^R_{Zb})^2\over (g^L_{Zb})^2+(g^R_{Zb})^2}~~ ,
\eq

the same quantity that is measured in the forward-backward $b\overline{b}$ asymmetry in $e^{+}e^{-}$
annihilation at the Z pole~\cite{LEPEWWG}:

\bq
A_{FB}^{b}= \frac{3}{4}\mathcal{A}_{e}\mathcal{A}_{b}~~ , ~~~~~~\mbox{where}~~~~~~\mathcal{A}_{e} = 
{(g^L_{Ze})^2-(g^R_{Ze})^2\over (g^L_{Ze})^2+(g^R_{Ze})^2}~~ .  
\eq

\begin{figure}
\vspace{1.5cm}
  \begin{center}
\resizebox{0.5\columnwidth}{!}{%
  \includegraphics{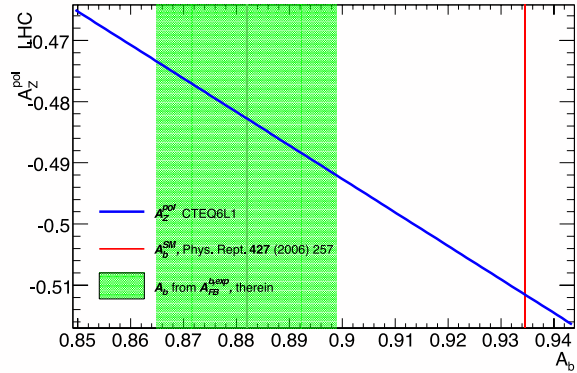} }
\caption{Polarization asymmetry $A_{Z}$ in bZ production at LHC with
  $\sqrt{s}=$7~TeV. The green band displays the $\pm 1\,\sigma$
  bounds~\cite{LEPEWWG}~for the measured asymmetry parameter
  $\mathcal{A}_{b}$ while the SM prediction~\cite{LEPEWWG}~is shown in red.}
\label{fig:Asym}       
\end{center}
\end{figure}

In order to exhibit the relation  between $A_{Z}$ and~$\mathcal{A}_{b}$
without any approximations, we have implemented a numerical calculation
of the full helicity
amplitude retaining the bottom mass
effects; in our calculation we require a final state b-quark with $p_{T}>$25 GeV
and rapidity $|y|<$2 to reproduce the typical
experimental phase space cuts.
The gluon and bottom quark in the initial state are folded
with CTEQ6~\cite{CTEQ} parton distribution functions.
The polarization asymmetry $A_{Z}$ in bZ
production at LHC with $\sqrt{s}=$7~TeV is shown in
Figure~\ref{fig:Asym} as function of $\mathcal{A}_{b}$~along with the
SM prediction~\cite{LEPEWWG}~(red band)~and
the measured LEP1 value~\cite{LEPEWWG}~(green band). 

As can be argued by inspection of Figure~\ref{fig:Asym}, the $A_{Z}$
measurement at the LHC could be sufficiently sensitive to
$\mathcal{A}_{b}$  in order to discriminate between LEP1 measurement and SM
prediction provided that a $\sim 8\%$ precision will be achieved on $A_{Z}$
measurement at LHC. To better realize if such required precision could be
reached in the  $A_{Z}$ calculation we need now to assess the effect of
its dominant theoretical uncertainties.

%
%
%


\subsection*{3. Impact of the scale/PDF choices and radiative corrections}

The previous discussion has been performed at the simplest Born level. 
The next relevant question is that of verifying whether the expression of $A_{Z}$ remains 
essentially identical when possible sources of theoretical uncertainties or NLO corrections are considered. 

We have proceeded in the following way. First, we have taken into
account possible effects of either strong scales or pdf variations;  
as shown in Ref.~\cite{ZbNloQCD}, these variations generate a sensible effect, of the almost ten percent relative size, in the total cross section. 
Next, we have considered the possible contribution of NLO electroweak
radiative corrections; their effect on the total and angular cross section have been determined in a recent paper~\cite{Denner}~ 
and found to be possibly relevant. 

\begin{figure}
\vspace{1.5cm}
  \begin{center}
\resizebox{0.5\columnwidth}{!}{%
  \includegraphics{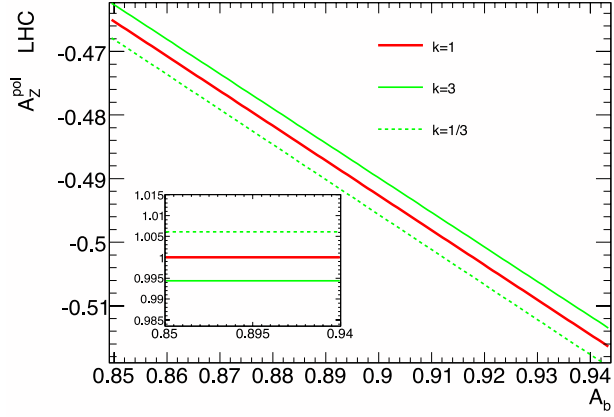} }
\caption{Polarization asymmetry $A_{Z}$ as function of
  $\mathcal{A}_{b}$ for three different choice of factorization and
  renormalization scales, respectively $\mu_{F}$ and $\mu_{R}$,
  $\mu_{F}=\mu_{R}=k\mu_{0}$ with $\mu_{0}=M_{Z}$ and $k=$1, 3 and 1/3. }
\label{fig:Scales}       
\end{center}
\end{figure}

The dependence of $A_{Z}$ on factorization and renormalization scales,
$\mu_{F}$ and $\mu_{R}$ respectively, is evaluated by varying their
values simultaneously by a conservative factor four with respect to
the central value; $A_{Z}$ is shown in
Figure~\ref{fig:Scales}~as function of $\mathcal{A}_{b}$ for
$\mu_{F}=\mu_{R}= k\mu_{0}$ with $\mu_{0}=M_{Z}$ and $k=$1, 3 and
1/3. As can be observed from Figure~\ref{fig:Scales}~effect of scales
variation on $A_{Z}$ is below $1\%$. However is worth noting that the
total cross section dependence on $\mu_{R}$ could be strongly
reduced by using the ``Principle of Maximum Conformality''
scale-setting (see for instance~\cite{Brod}).    

\begin{figure}
\vspace{1.5cm}
  \begin{center}
\resizebox{0.5\columnwidth}{!}{%
  \includegraphics{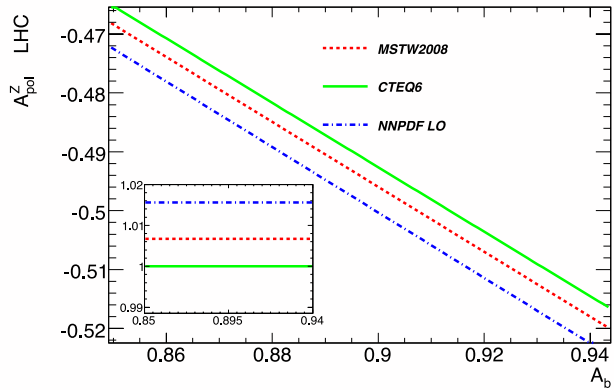} }
\caption{Polarization asymmetry $A_{Z}$  as function of
  $\mathcal{A}_{b}$ for three different choice pdf set as described in the text. }
\label{fig:Pdf}       
\end{center}
\end{figure}

The asymmetry dependence on the pdf is examined performing the
numerical calculation with different pdf sets. In Figure~\ref{fig:Pdf}, we present $A_{Z}$ as function of
$\mathcal{A}_{b}$ for three different LO pdf sets:
CTEQ~\cite{CTEQ}; MSTW2008~\cite{MSTW}~and
NNPDF~\cite{NNPDF}. As one sees, the dependence on the pdf set is
below $2\%$ while the total cross section can be affected by large variations of order $7\%$~\cite{ATLASZb}.

The NLO EW effects on $A_{Z}$ deserve a rather different discussion. In principle, these
effects would not introduce any appreciable theoretical uncertainty, since the values of the
involved parameters are all known with great accuracy. The goal of their calculation would 
simply be that of offering a more complete theoretical prediction for $A_{Z}$. In fact,
it is well known that electroweak corrections can have sizable
effects on processes involving W or Z production at LHC. We
have observed it in associated top and W production~\cite{tW1,tW2}~and recent
papers on W+jet or Z+jet production had also mentioned it,
see~\cite{Denner,Wjets}. These effects can reach the several percent size and
even more than ten percent on the subprocess cross sections. This can
be immediately understood by looking at the simple Sudakov (squared
and linear) logarithmic terms which affect the amplitudes at high
energy~\cite{HighELog1,HighELog2}. To estimate the size of this type of
effect at lower energies one also can use the so-called ``augmented Sudakov'' terms, in which constant terms have been
added to the logarithmic ones~\cite{AugSud}.   
Using this approach, one can immediately be convinced that the polarization asymmetry $A_{Z}$
will be essentially not affected by these electroweak corrections.
Actually, looking at the transverse Born amplitudes, one first remarks  that
because $g^L_{Zb} \sim 5 \, g^R_{Zb}$, the dominant amplitudes
are $F_{-+-+}$ and $F_{----}$. The other ones will contribute to the
total cross section by terms suppressed by a factor $1/25$.
Then, applying the Sudakov rules of
Ref.~\cite{tW1,tW2,HighELog1,HighELog2,AugSud}, one sees that 
the leading logs associated to the $b_L$ and Z states
are very similar for these two amplitudes. A raw estimate gives 
effects of several percents in the 1 TeV range which should directly affect the cross section.

However for $A_Z$, dominated by
$(|F_{-+-+}|^2 - |F_{----}|^2)/( |F_{-+-+}|^2 + |F_{----}|^2 )$ ratio,
the common electroweak corrections to each of these
amplitudes in the numerator and in the denominator
will cancel out.
So a small non zero effect will only come from the smaller amplitudes 
(which contribute by a factor $25$ less) and from the small differences due to the 
subleading (mass suppressed) terms.

Using the augmented Sudakov expressions written in ref.~\cite{AugSud}
we have checked that the effects on $A_{Z}$ reach at most the one percent level.
In this spirit, we shall consider in this preliminary paper the SM NLO electroweak
corrections as probably irrelevant. A more complete determination of their numerical
 effect will be given in a forthcoming paper.

\subsection*{4. Conclusion} 

We have shown that the Z polarization asymmetry $A_{Z}$ in 
bZ production at the LHC is strictly connected to the well known 
forward-backward $b\overline{b}$ asymmetry at Z pole, $A^{b}_{FB}$, measured at LEP1.
Our results indicate that $A_{Z}$ is almost free from
theoretical uncertainties related to QCD scale variations  as well as to
pdf set variations; this property strongly suggests in our opinion a measurement of $A_{Z}$ at LHC
as a unique candidate to possibly clarify 
 the long standing puzzle related to the
$A^{b}_{FB}$ measurement at LEP1.

More in general it can be observed that polarization asymmetry
observables would be quite relevant
theoretical observables at LHC, as shown by a recent paper~\cite{tH-},
where a polarization asymmetry was studied in the context of polarized
top production in association with a charged Higgs boson, as a possible
way of determining the $\tan\beta$ parameter in the MSSM.
 A rather general conclusion of our paper is therefore in our opinion  that measurements of 
polarization at LHC would represent a tough but possibly quite rewarding experimental effort.
A more complete discussion of  Z polarization measurements at LHC will be treated in a 
forthcoming paper.

{\bf Acknowledgements.}
We thank M. Cobal and M. Pinamonti for several discussions concerning the measurement of 
polarization at LHC.


\end{document}